# Second order gradient ascent pulse engineering

P. de Fouquieres[1], S.G. Schirmer[1,*], S.J. Glaser[2,*], Ilya Kuprov[3,*]

[1]*Centre for Quantum Computation,
Department of Applied Mathematics and Theoretical Physics,
University of Cambridge, Wilberforce Road,
Cambridge CB3 0WA, United Kingdom.*

[2]*Department of Chemistry, Technische Universität München,
85747 Garching, Germany.*

[3]*Oxford e-Research Centre, University of Oxford,
7 Keble Road, Oxford OX1 3QG, United Kingdom.*

Email addresses:
sgs29@cam.ac.uk       (SGS)
glaser@tum.de         (SJG)
ilya.kuprov@oerc.ox.ac.uk   (IK)

**Abstract**

We report some improvements to the gradient ascent pulse engineering (GRAPE) algorithm for optimal control of spin ensembles and other quantum systems. These include more accurate gradients, convergence acceleration using the Broyden-Fletcher-Goldfarb-Shanno (BFGS) quasi-Newton algorithm as well as faster control derivative calculation algorithms. In all test systems, the wall clock time and the convergence rates show a considerable improvement over the approximate gradient ascent.

# 1. Introduction

An *optimal control* problem consists in bringing a dynamic system from one state to another to a given accuracy with minimum expenditure of effort [1-3]. Such tasks are encountered in optical spectroscopy [4-6], magnetic resonance [7-12], spin dynamics [13,14] and the emerging field of quantum information processing [15-19]. While many variations exist in practice [20-22], depending on the desired outcome and the constraints placed on the solution by instrumental limitations [1,2,23], they can all be broadly classified into *gate design* problems [11,24], where a specific unitary transformation of the entire state space is sought, and *point-to-point* transfer problems [10,25-27], where the population is to be moved from one specific state to another without conditions on the dynamics of other states. Because any gate design problem can be represented as a point-to-point transfer problem in a space of higher dimension [23,24], we will only consider the point-to-point formulation below.

The state of a quantum system can be described by a density operator $\hat{\rho}(t)$, whose evolution is governed by the quantum Liouville equation [28]:

$$\frac{\partial}{\partial t}\hat{\rho}(t) = -\mathrm{i}[\hat{H}(t), \hat{\rho}(t)] + \hat{\hat{R}}\hat{\rho}(t) \tag{1}$$

where $\hat{H}(t)$ is a possibly-time dependent Hamiltonian and $\hat{\hat{R}}$ is the relaxation superoperator. It is often convenient to carry out the calculations in Liouville space by replacing a matrix representation of $\hat{\rho}(t)$ with a vector $|\hat{\rho}(t)\rangle$ obtained by stacking the columns of $\hat{\rho}(t)$. In this representation, the equation acquires the following form:

$$\frac{\partial}{\partial t}|\hat{\rho}(t)\rangle = -\mathrm{i}\hat{\hat{L}}(t)|\hat{\rho}(t)\rangle, \qquad \hat{\hat{L}}(t) = \hat{E} \otimes \hat{H}(t) - \hat{H}(t)^{\mathrm{T}} \otimes \hat{E} + \mathrm{i}\hat{\hat{R}} \tag{2}$$

where $\hat{E}$ is the unit matrix of the same dimension as $\hat{H}$ [28] and $\hat{\hat{R}}$ is the Liouville space representation of the relaxation superoperator. The general solution may be formally written as:

$$|\hat{\rho}(t)\rangle = \exp_{(\mathrm{O})}\left(-\mathrm{i}\int_0^t \hat{\hat{L}}(t)\,dt\right)|\hat{\rho}(t)\rangle \tag{3}$$

where $\exp_{(\mathrm{O})}$ indicates Dyson's time-ordered exponential [29]. Given a fixed grid of points $\{t_1,...,t_N\}$, the density matrix at a particular grid point $n$ is then given by:

$$|\hat{\rho}(t_n)\rangle = \hat{\hat{P}}_n...\hat{\hat{P}}_2\hat{\hat{P}}_1|\hat{\rho}(0)\rangle \qquad \hat{\hat{P}}_n = \exp_{(\mathrm{O})}\left(-\mathrm{i}\int_{t_{n-1}}^{t_n}\hat{\hat{L}}(t)\,dt\right) \tag{4}$$

A point-to-point transfer problem consists in finding such $\hat{H}(t)$ as would maximize the population of a given target density matrix $|\hat{\sigma}\rangle$ after evolution from a given initial state $|\hat{\rho}(0)\rangle$ under the total Liouvillian $\hat{\hat{L}}(t)$ [23,27]:

$$\hat{H}_{\text{opt}}(t) \in \arg\max_{\hat{H}(t)} \left( \langle \hat{\sigma} | \hat{\rho}(t_N) \rangle \right) \tag{5}$$

where the maximum is sought in the class of Hermitian matrix valued functions of time. From the experimental perspective, not every part of $\hat{H}(t)$ can be modified at will, and it is common to separate it into the "drift" and the "control" parts:

$$\hat{H}(t) = \hat{H}_0 + \sum_k c^{(k)}(t) \hat{H}_k \quad \Rightarrow \quad \hat{\hat{L}}(t) = \hat{\hat{L}}_0 + \sum_k c^{(k)}(t) \hat{\hat{L}}_k$$

$$\hat{\hat{L}}_0 = \hat{E} \otimes \hat{H}_0 - \hat{H}_0^{\text{T}} \otimes \hat{E} + i\hat{\hat{R}}, \quad \hat{\hat{L}}_k = \hat{E} \otimes \hat{H}_k - \hat{H}_k^{\text{T}} \otimes \hat{E} \tag{6}$$

$\hat{H}_0$ being the "drift" component deemed to be beyond our direct influence and $\hat{H}_k$ are the "control" components, whose contributions may be varied experimentally [13]. Various constraints are often placed on the control functions $c^{(k)}(t)$, mostly to enforce the instrumental limitations [20,21]. The optimization problem in Equation (5) is difficult to solve in full generality, and it is common to simplify the description of $\hat{H}(t)$ by assuming the control functions to be piecewise-constant [9,12]:

$$c^{(k)}(t) = c_n^{(k)} \qquad t_{n-1} < t < t_n \tag{7}$$

In practice, this is often not an approximation, since the actual output of many hardware devices, *e.g.* waveform generators in NMR spectroscopy, can be made piecewise-constant. Under this assumption, the time-ordered exponential (a notoriously complicated object from the numerical calculation perspective) in Equation (4) simplifies into a simple matrix exponential:

$$\hat{\hat{P}}_n = \exp_{(O)}\left[ -i \int_{t_{n-1}}^{t_n} \hat{\hat{L}}(t) dt \right] = \exp\left[ -i \left( \hat{\hat{L}}_0 + \sum_k c_n^{(k)} \hat{\hat{L}}_k \right) \Delta t \right] \tag{8}$$

where $\Delta t$ is the time grid spacing. Progress can then be made with the optimization problem in Equation (5), because the gradient of the error functional with respect to the amplitude of control $k$ at time step $n$ is now easily computed:

$$\frac{\partial}{\partial c_n^{(k)}} \langle \hat{\sigma} | \hat{\rho}(t_N) \rangle = \frac{\partial}{\partial c_n^{(k)}} \langle \hat{\sigma} | \hat{\hat{P}}_N \ldots \hat{\hat{P}}_n \ldots \hat{\hat{P}}_1 | \hat{\rho}(0) \rangle = \langle \hat{\sigma} | \hat{\hat{P}}_N \ldots \hat{\hat{P}}_{n+1} \frac{\partial \hat{\hat{P}}_n}{\partial c_n^{(k)}} \hat{\hat{P}}_{n-1} \ldots \hat{\hat{P}}_1 | \hat{\rho}(0) \rangle =$$

$$= \left( \hat{\hat{P}}_{n+1}^\dagger \ldots \hat{\hat{P}}_N^\dagger | \hat{\sigma} \rangle \right)^\dagger \frac{\partial \hat{\hat{P}}_n}{\partial c_n^{(k)}} \left( \hat{\hat{P}}_{n-1} \ldots \hat{\hat{P}}_1 | \hat{\rho}(0) \rangle \right) \tag{9}$$

This effectively means that the destination state has to be propagated *backward* to time point $t_n$, the source state has to be propagated *forward* to time point $t_{n-1}$ and the scalar product with the derivative of the propagator for the time step $n$ has to be taken. In practice [7,9,12,27], the entire forward trajectory is computed from $\hat{\rho}(0)$, the entire backward trajectory is computed from $\hat{\sigma}$ and then the two are folded in each step with the propagator derivatives as prescribed by Equation (9).

## 2. Calculation of control derivatives

The expression for the propagator derivative suggested in the paper introducing the GRAPE method [9] reasonably assumes that the control sequence discretization step is small:

$$\frac{\partial}{\partial c_n^{(k)}} \hat{P}_n = \frac{\partial}{\partial c_n^{(k)}} \exp\left[-\mathrm{i}\left(\hat{L}_0 + \sum_k c_n^{(k)} \hat{L}_k\right)\Delta t\right] = $$
$$= \exp\left[-\mathrm{i}\left(\hat{L}_0 + \sum_k c_n^{(k)} \hat{L}_k\right)\Delta t\right]\left(-\mathrm{i}\hat{L}_k \Delta t + O(\Delta t^2)\right) = \hat{P}_n\left(-\mathrm{i}\hat{L}_k \Delta t\right) + O(\Delta t^2) \quad (10)$$

this assumption makes the evaluation of control gradient very computationally affordable – it introduces no new matrix-matrix multiplications beyond those used to compute the propagators, because $-\mathrm{i}\hat{L}_k \Delta t$ and $\hat{P}_n$ can be multiplied sequentially into the vectors on either side of the derivative in Equation (9). The cost of the control gradient is therefore approximately equal to the cost of the trajectory calculation. From Equation (9), we have:

$$\frac{\partial}{\partial c_n^{(k)}} \langle\hat{\sigma}|\hat{\rho}(t_N)\rangle = \left(\hat{P}_n^\dagger \ldots \hat{P}_N^\dagger |\hat{\sigma}\rangle\right)^\dagger \left(-\mathrm{i}\hat{L}_k \Delta t\right)\left(\hat{P}_{n-1}\ldots\hat{P}_1|\hat{\rho}(0)\rangle\right) + O(\Delta t^2) \quad (11)$$

While the gradient ascent using this equation does in most cases yield acceptable accuracy solutions, it has been recognized for some time that the $O(\Delta t^2)$ residual tends to limit both the convergence rate and the final accuracy achievable. As the gradient becomes smaller during the minimization, it is the term under the trace in Equation (11) that gets reduced, and the situation eventually emerges where the approximation error dominates the gradient. This leads to the often observed and much lamented "slowdown" of the GRAPE algorithm as it approaches high transfer fidelities. It also scrambles the approximate Hessians used by quasi-Newton methods, essentially preventing their use. This may be seen directly by following the $O(\Delta t^2)$ residual through Newton's method:

$$f(\vec{x} + \Delta\vec{x}) = f(\vec{x}) + \left(\nabla f(\vec{x})^\mathrm{T} + O(\Delta t^2)\right)\Delta\vec{x} + \frac{1}{2}\Delta\vec{x}\mathbf{H}\Delta\vec{x} + O(|\Delta\vec{x}|^3)$$
$$= f(\vec{x}) + \nabla f(\vec{x})^\mathrm{T} \Delta\vec{x} + \frac{1}{2}\Delta\vec{x}\mathbf{H}\Delta\vec{x} + O(\Delta t^2 |\Delta\vec{x}|) + O(|\Delta\vec{x}|^3) \quad (12)$$

Unless the time step $\Delta t$ is chosen to be extremely small, the $O(\Delta t^2 |\Delta \vec{x}|)$ error term, which is linear in $|\Delta \vec{x}|$, completely obscures the Hessian term, which is quadratic in $|\Delta \vec{x}|$. For typical NMR systems, this accuracy constraint places $\Delta t$ into nanosecond range, which makes the number of steps very large and causes difficulties on the instrumental side.

In our experience, these problems can be removed at a reasonable computational cost, if the *exact* propagator gradient is used, rather than the first order approximation. The most straightforward avenue is to differentiate the Taylor or Chebyshev expansion for the exponential directly [30]. In the case of the Taylor series, this yields:

$$\frac{\partial}{\partial c_n^{(k)}} \exp\left[-\mathrm{i}\hat{\hat{L}}\Delta t\right] = \sum_{p=1}^{\infty} \frac{(-\mathrm{i}\Delta t)^p}{p!} \sum_{q=0}^{p-1} \hat{\hat{L}}^q \hat{\hat{L}}_k \hat{\hat{L}}^{p-q-1}, \qquad \hat{\hat{L}} = \hat{\hat{L}}_0 + \sum_k c_n^{(k)} \hat{\hat{L}}_k \tag{13}$$

The second sum appears because $\hat{\hat{L}}$ and $\hat{\hat{L}}_k$ do not necessarily commute. This formulation is computationally about as expensive as the original exponential because matrices involved (particularly $\hat{\hat{L}}_k$) are often very sparse [31,32], but it is rather inconvenient because it involves double summation. A more computer-friendly version is given by a commutator series, which is the direct extension of Equation (10):

$$\frac{\partial}{\partial c_n^{(k)}} \hat{\hat{P}}_n = \frac{\partial}{\partial c_n^{(k)}} \exp\left[-\mathrm{i}\hat{\hat{L}}\Delta t\right] = $$
$$= \exp\left[-\mathrm{i}\hat{\hat{L}}\Delta t\right]\left(-\mathrm{i}\hat{\hat{L}}_k \Delta t + \frac{\Delta t^2}{2}[\hat{\hat{L}},\hat{\hat{L}}_k] + \frac{\mathrm{i}\Delta t^3}{6}[\hat{\hat{L}},[\hat{\hat{L}},\hat{\hat{L}}_k]] - \frac{\Delta t^4}{24}[\hat{\hat{L}},[\hat{\hat{L}},[\hat{\hat{L}},\hat{\hat{L}}_k]]] + \ldots\right) \tag{14}$$

This expression can be obtained by rotating summation indices in Equation (13):

$$\sum_{p=1}^{\infty} \frac{(-\mathrm{i}\Delta t)^p}{p!} \sum_{q=0}^{p-1} \hat{\hat{L}}^q \hat{\hat{L}}_k \hat{\hat{L}}^{p-q-1} = \sum_{p=0}^{\infty}\sum_{q=0}^{\infty} \frac{A^p B A^q}{(p+q+1)!}, \qquad A = -\mathrm{i}\hat{\hat{L}}\Delta t, \quad B = -\mathrm{i}\hat{\hat{L}}_k \Delta t, \tag{15}$$

splitting the factorial in the denominator, then summing the series into matrix exponentials

$$\frac{1}{(p+q+1)!} = \frac{1}{p!q!}\int_0^1 (1-\alpha)^p \alpha^q d\alpha \quad \Rightarrow \quad \sum_{p=0}^{\infty}\sum_{q=0}^{\infty} \frac{A^p B A^q}{(p+q+1)!} = e^A \int_0^1 e^{-\alpha A} B e^{\alpha A} d\alpha, \tag{16}$$

evaluating the integral

$$\int_0^1 e^{-\alpha A} B e^{\alpha A} d\alpha = \int_0^1 e^{-\mathrm{ad}_A \alpha} B d\alpha = \gamma[-\mathrm{ad}_A] B,$$
$$\gamma(z) = \frac{e^z - 1}{z} = \sum_{n=0}^{\infty} \frac{z^n}{(n+1)!}, \qquad \mathrm{ad}_x y = [x, y] \tag{17}$$

and expressing powers of $\text{ad}_A$ as nested commutators with $A$:

$$\sum_{m=0}^{\infty}\frac{(-\text{ad}_A)^m}{(m+1)!}B = \sum_{m=0}^{\infty}\frac{(-1)^m}{(m+1)!}[A,B]_m, \qquad [A,B]_m = [A,[A,B]_{m-1}], \qquad [A,B]_0 = B \qquad (18)$$

With Equations (14)-(18) in place, the expression for the control gradient becomes:

$$\frac{\partial}{\partial c_n^{(k)}}\langle\hat{\sigma}|\hat{\rho}(t_N)\rangle = -\left(\hat{\hat{P}}_n^\dagger\ldots\hat{\hat{P}}_N^\dagger|\hat{\sigma}\rangle\right)^\dagger\left(\sum_{m=0}^{\infty}\frac{(\text{i}\Delta t)^{m+1}}{(m+1)!}[\hat{\hat{L}}_0 + \sum_k c_n^{(k)}\hat{\hat{L}}_k, \hat{\hat{L}}_k]_m\right)\left(\hat{\hat{P}}_{n-1}\ldots\hat{\hat{P}}_1|\hat{\rho}(0)\rangle\right), \qquad (19)$$

where the summation of the series is to be continued until the desired accuracy (as indicated by the residual norm) is achieved. A few initial orders of accuracy in the Taylor expansion of $\gamma(z)$ are plotted in Figure 1.

The numerical accuracy of Equation (19) in finite-precision arithmetic merits further discussion. As with all power series, the perfect scenario from the numerical point of view is to have the norm of the argument $-\text{i}\hat{\hat{L}}\Delta t$ scaled into the unit interval – this avoids the "hump" in the convergence and keeps the terms well within the accuracy limits imposed by 64-bit floating-point arithmetic. As Figure 2 demonstrates, adequate numerical accuracy is maintained for $\|-\text{i}\hat{\hat{L}}\Delta t\| < 30$, but deteriorates rapidly thereafter. The standard technique used to resolve this issue is known as "scaling and squaring" [30,33]:

$$\exp\left(-\text{i}\hat{\hat{L}}\Delta t\right) = \exp\left(\frac{-\text{i}\hat{\hat{L}}\Delta t}{2}\right)^2. \qquad (20)$$

The product rule for the derivative makes the scaling and squaring procedure for the derivative propagator slightly different:

$$\frac{\partial}{\partial c_n^{(k)}}\exp\left(-\text{i}\hat{\hat{L}}\Delta t\right) = \exp\left(\frac{-\text{i}\hat{\hat{L}}\Delta t}{2}\right)\left[\frac{\partial}{\partial c_n^{(k)}}\exp\left(\frac{-\text{i}\hat{\hat{L}}\Delta t}{2}\right)\right] + \left[\frac{\partial}{\partial c_n^{(k)}}\exp\left(\frac{-\text{i}\hat{\hat{L}}\Delta t}{2}\right)\right]\exp\left(\frac{-\text{i}\hat{\hat{L}}\Delta t}{2}\right) \qquad (21)$$

Because the exponential propagator itself is computed elsewhere in the GRAPE procedure [9], the cost of the squaring step is modest – two sparse matrix multiplications. Our experience with this procedure has been very positive, it tolerates scaling factors in excess of $10^6$, thus encompassing all practically encountered GRAPE algorithm application situations.

If relaxation is negligible and the Hamiltonian is small enough to be diagonalized, the series in Equation (19) may be avoided because we can evaluate $\gamma[\text{ad}(\text{i}\hat{H}\Delta t)](-\text{i}\hat{H}_k\Delta t)$ directly by diagonalizing $\hat{H}$ [8,22,30]. Let $\hat{H} = \hat{V}\hat{\Lambda}\hat{V}^\dagger$, where $\hat{V}$ is a unitary matrix whose columns are ei-genvectors of $\hat{H}$ and $\hat{\Lambda}$ is a diagonal matrix with the corresponding eigenvalues $\lambda_r$ along the diagonal. We then have:

$$\hat{D}^{(k)} = \gamma\left[\text{ad}\left(i\hat{H}\Delta t\right)\right]\left(-i\hat{H}_k\Delta t\right) = \hat{V}\left(\hat{G}\odot\hat{B}\right)\hat{V}^\dagger$$
$$G_{rs} = \gamma\left[i(\lambda_r - \lambda_s)\Delta t\right], \qquad \hat{B} = \hat{V}^\dagger\left(-i\hat{H}_k\Delta t\right)\hat{V}$$
(22)

where $\odot$ denotes element-wise (Hadamard) matrix multiplication. Using this formula, we have:

$$\frac{\partial}{\partial c_n^{(k)}}\langle\hat{\sigma}|\hat{\rho}(t_N)\rangle = \left(\hat{P}_n^\dagger\ldots\hat{P}_N^\dagger|\hat{\sigma}\rangle\right)^\dagger \hat{\hat{D}}^{(k)}\left(\hat{P}_{n-1}\ldots\hat{P}_1|\hat{\rho}(0)\rangle\right)$$
$$\hat{\hat{D}}^{(k)} = \hat{D}^{(k)}\otimes\hat{E} - \hat{E}\otimes\hat{D}^{(k)\text{T}}$$
(23)

This method is the adaptation of the diagonalization method for matrix functions – the matrix is transformed into its eigenframe, the function is applied to the eigenvalues and the result is transformed back into the original frame. It is not applicable to Hamiltonians with dimensions in excess of $10^4$, because the eigenvector array $\hat{V}$ is often dense even for sparse Hamiltonians and overflows the computer memory.

Equations (13)-(23) present an analytical formalism for the calculation of the control derivatives. A popular numerical alternative is to use finite-difference approximations, *e.g.*:

$$\frac{\partial\hat{\hat{P}}_n}{\partial c_n^{(k)}} = \frac{\hat{\hat{P}}_n\left(\ldots,c_n^{(k)}+h,\ldots\right) - \hat{\hat{P}}_n\left(\ldots,c_n^{(k)},\ldots\right)}{h} + O(h)$$
(24)

$$\frac{\partial\hat{\hat{P}}_n}{\partial c_n^{(k)}} = \frac{\hat{\hat{P}}_n\left(\ldots,c_n^{(k)}+h,\ldots\right) - \hat{\hat{P}}_n\left(\ldots,c_n^{(k)}-h,\ldots\right)}{2h} + O(h^2)$$
(25)

where the amplitude of the $k$-th control at the $n$-th time point is varied by a finite amount $h$. Equations (24) and (25) are indicative – they are the simplest examples of a large class of numerical finite-difference approximations for the derivative [34]. The primary balance to be maintained in this approach is between the approximation accuracy, the numerical accuracy and the computational cost of the derivative.

The forward finite difference approximation in Equation (24) has the advantage of being computationally affordable – it only requires the calculation of one extra $\exp(-i\hat{\hat{L}}\Delta t)\hat{\rho}$ product per step, which may be carried out using Krylov subspace techniques, thus avoiding matrix multiplications. From Equation (9):

$$\frac{\partial}{\partial c_n^{(k)}}\left\langle\hat{\sigma}\left|\hat{\rho}(t_N)\right.\right\rangle = \left(\hat{\hat{P}}_{n+1}^{\dagger}...\hat{\hat{P}}_N^{\dagger}|\sigma\rangle\right)\frac{\partial \hat{\hat{P}}_n}{\partial c_n^{(k)}}\left(\hat{\hat{P}}_{n-1}...\hat{\hat{P}}_1|\hat{\rho}(0)\rangle\right) =$$
$$= \left(\hat{\hat{P}}_{n+1}^{\dagger}...\hat{\hat{P}}_N^{\dagger}|\sigma\rangle\right)\frac{\hat{\hat{P}}_n\left(c_n^{(k)}+h\right)-\hat{\hat{P}}_n\left(c_n^{(k)}\right)}{h}\left(\hat{\hat{P}}_{n-1}...\hat{\hat{P}}_1|\hat{\rho}(0)\rangle\right) + O(h)$$
(26)

Approximations with a higher order of accuracy may be used at the expense of having to calculate further $\exp(-i\hat{\hat{L}}\Delta t)\hat{\rho}$ products for the extra stencil points. In common with the commutator series approach in Equation (19), the finite difference method is applicable to dissipative quantum systems, where anti-Hermitian terms may be present in the Liouvillian. It should be noted that Equation (26) only involves a finite difference with respect to the step propagator – the rest of the trajectory does not need to be recomputed. This is a much more efficient arrangement as compared to the brute-force finite-differencing of $\langle\hat{\sigma}|\hat{\rho}(t_N)\rangle$.

The accuracy of finite difference methods depends on the step size $h$. In practical situations, the choice is constrained in two ways: if the step is too large, the finite difference would not be a good approximation to the derivative, and if the step is too small, the number of accurate digits in the floating point representation of the difference would reduce to none. In general, we do not have sufficient information to make an *a priori* estimate for the finite difference approximation error (it requires the knowledge of higher derivatives), but the numerical round-off error is a somewhat more straightforward quantity. A reasonable strategy therefore is to choose the smallest $h$ for which the round-off error is guaranteed to be below a given threshold. Assuming the approximation error is indeed small for that choice of $h$, we can approximate a function $f(x)$ with a linear polynomial $f(x+h) = f(x) + f'(x)h$ for the purpose of obtaining the required round-off error bound.

The evaluation of matrix exponentials is accurate up to a fixed purely absolute error $\varepsilon_A$, which is a few orders of magnitude larger than the machine precision $\varepsilon_M$ (equal to $2.22\cdot10^{-16}$ in 64-bit arithmetic), because the norm of exponential time propagators in quantum mechanics is less than or equal 1. The error incurred in computing $f(x+h) - f(x)$ is then $2\varepsilon_A$ plus at most $\varepsilon_M(|f(x)|+|f'(x)|h)$ from truncating their difference. The finite difference approximation $|h^{-1}[f(x+h)-f(x)] - f'(x)|$ then carries an absolute error bounded by:

$$\frac{1}{|h|}\left(2\varepsilon_A + \varepsilon_M|f(x)|\right) + \varepsilon_M|f'(x)|. \tag{27}$$

This expression may be equated to the chosen error threshold and solved for $h$, assuming that an order of magnitude estimate of $|f'(x)|$ is available. Even if the norm of $f'(x)$ cannot be estimated *a priori*, Equation (27) can still be used to validate the choice of step *a posteriori*, using the finite difference approximation to $f'(x)$.

The dependence of the approximation accuracy on the finite difference step size is illustrated in Figure 3 – for large steps the error is dominated by the approximation error of the finite difference, which drops smoothly when the step is reduced. For small steps the error is dominated by the numerical round-off errors, which increase erratically as the numerical accuracy decreases.

The choice of the differentiation algorithm is ultimately left to user's discretion. The considerable improvement that better gradient accuracy brings to the asymptotic convergence rate is illustrated in Figure 4 – for the spin chain in question, a pulse with 100 nanosecond time stepping is clearly outside the validity range of the first-order approximation in Equation (11), and further terms in Equation (19) are necessary.

## 3. GRAPE with quasi-Newton optimizers

Second-order optimal control algorithms are well researched (see, *e.g.* [35-37]) and have been successfully applied in areas outside of magnetic resonance – for example in power control [38] and fluid flow optimization [39]. In the magnetic resonance context, the above noted fact that the control gradient of the objective function is relatively cheap to compute means that it is almost always advantageous to use the gradient history to build an approximation to the Hessian matrix, which can then be used in quasi-Newton optimization algorithms, which can exhibit super-linear convergence [40]. Because GRAPE is a concurrent update algorithm [9], the standard quasi-Newton methods may be used directly.

Several schemes exist for generating approximate Hessians from the gradient history, the most notable being DFP [41] and BFGS [40]:

$$\mathbf{H}_{k+1}^{\text{DFP}} = \left( \mathbf{E} - \frac{\vec{g}_k \vec{s}_k^{\text{T}}}{\vec{g}_k^{\text{T}} \vec{s}_k} \right) \mathbf{H}_k \left( \mathbf{E} - \frac{\vec{s}_k \vec{g}_k^{\text{T}}}{\vec{g}_k^{\text{T}} \vec{s}_k} \right) + \frac{\vec{g}_k \vec{g}_k^{\text{T}}}{\vec{g}_k^{\text{T}} \vec{s}_k}$$

$$\mathbf{H}_{k+1}^{\text{BFGS}} = \mathbf{H}_k + \frac{\vec{g}_k \vec{g}_k^{\text{T}}}{\vec{g}_k^{\text{T}} \vec{s}_k} - \frac{(\mathbf{H}_k \vec{s}_k)(\mathbf{H}_k \vec{s}_k)^{\text{T}}}{\vec{s}_k^{\text{T}} \mathbf{H}_k \vec{s}_k} \qquad (28)$$

$$\vec{g}_k = \nabla f(\vec{x}_{k+1}) - \nabla f(\vec{x}_k), \qquad \vec{s}_k = \vec{x}_{k+1} - \vec{x}_k$$

These pseudo-Hessians are constructed to satisfy the natural finite difference condition:

$$\nabla f(\vec{x}_k) = \nabla f(\vec{x}_{k+1}) - \mathbf{H}_{k+1}(\vec{x}_{k+1} - \vec{x}_k). \tag{29}$$

A necessary condition for $\mathbf{H}_{k+1}$ to be negative definite is therefore that

$$(\vec{x}_{k+1} - \vec{x}_k)^{\mathrm{T}} \mathbf{H}_{k+1}(\vec{x}_{k+1} - \vec{x}_k) = (\vec{x}_{k+1} - \vec{x}_k)^{\mathrm{T}} (\nabla f(\vec{x}_{k+1}) - \nabla f(\vec{x}_k)) < 0, \tag{30}$$

but a useful property of BFGS and DFP update rules is that it is also a sufficient condition, assuming that $\mathbf{H}_0$ was chosen to be negative definite [40].

Because matrix inversions are expensive, it is in practice necessary to use the corresponding update schemes for the inverse of the Hessian:

$$\begin{aligned}
\left(\mathbf{H}^{-1}\right)_{k+1}^{\mathrm{BFGS}} &= \left(\mathbf{E} - \frac{\vec{g}_k \vec{s}_k^{\mathrm{T}}}{\vec{g}_k^{\mathrm{T}} \vec{s}_k}\right)^{\mathrm{T}} \mathbf{H}_k^{-1} \left(\mathbf{E} - \frac{\vec{g}_k \vec{s}_k^{\mathrm{T}}}{\vec{g}_k^{\mathrm{T}} \vec{s}_k}\right) + \frac{\vec{s}_k \vec{s}_k^{\mathrm{T}}}{\vec{g}_k^{\mathrm{T}} \vec{s}_k} \\
\left(\mathbf{H}^{-1}\right)_{k+1}^{\mathrm{DFP}} &= \mathbf{H}_k^{-1} + \frac{\vec{s}_k \vec{s}_k^{\mathrm{T}}}{\vec{g}_k^{\mathrm{T}} \vec{s}_k} - \frac{(\mathbf{H}_k^{-1} \vec{g}_k)(\mathbf{H}_k^{-1} \vec{g}_k)^{\mathrm{T}}}{\vec{g}_k^{\mathrm{T}} \mathbf{H}_k^{-1} \vec{g}_k}
\end{aligned} \tag{31}$$

In the case of BFGS, a very memory-efficient procedure is available for generating the next step vector directly from the past gradient history, requiring no matrix storage. It is known as memory-limited BFGS, or L-BFGS [42,43]. In the context of optimal control, the number of variables often exceeds $10^4$, and L-BFGS is the only quasi-Newton method that is capable of handling such problems. The performance of DFP, BFGS and L-BFGS for the optimization of a broadband magnetization inversion pulse in NMR spectroscopy is illustrated in Figure 5.

## 5. Conclusions and outlook

The GRAPE algorithm for control sequence optimization has the benefit of computationally affordable gradients. Using the equations reported in this paper, their accuracy may be improved beyond the first order approximation and the result used to generate approximate Hessians for quasi-Newton optimization. In all test systems, the wall clock time and the convergence rates show a considerable improvement over the approximate gradient ascent – the "slowdown" problem disappears. The BFGS-GRAPE procedure reported in this paper is implemented in the *Spinach* software library [32].

**Acknowledgements**

The authors would like to thank Burkhard Luy, Shai Machnes, Ivan Maximov, Uwe Sander, Thomas Schulte-Herbrüggen, Thomas Skinner, Luke Edwards and David Tannor for stimulating discussions. This work is supported by EPSRC (EP/F065205/1, EP/H003789/1,

<div style="text-align:center">References</div>

[1]     F.L. Lewis, V.L. Syrmos, Optimal control, Wiley, 1995.
[2]     L.S. Pontryagin, Mathematical Theory of Optimal Processes, Pergamon Press, 1964.
[3]     V.F. Krotov, Global methods in optimal control theory, Marcel Dekker, New York, 1996.
[4]     C.P. Koch, M. Ndong, R. Kosloff, Two-photon coherent control of femtosecond photoassociation, Faraday Disc., 142 (2009) 389-402.
[5]     C.P. Koch, J.P. Palao, R. Kosloff, F. Masnou-Seeuws, Stabilization of ultracold molecules using optimal control theory, Phys. Rev. A, 70 (2004) 013402.
[6]     J.P. Palao, R. Kosloff, C.P. Koch, Protecting coherence in optimal control theory: State-dependent constraint approach, Phys. Rev. A, 77 (2008).
[7]     N.I. Gershenzon, K. Kobzar, B. Luy, S.J. Glaser, T.E. Skinner, Optimal control design of excitation pulses that accommodate relaxation, J. Magn. Reson., 188 (2007) 330-336.
[8]     T.O. Levante, T. Bremi, R.R. Ernst, Pulse-sequence optimization with analytical derivatives. Application to deuterium decoupling in oriented phases, J. Magn. Reson., 121 (1996) 167-177.
[9]     N. Khaneja, T. Reiss, C. Kehlet, T. Schulte-Herbrüggen, S.J. Glaser, Optimal control of coupled spin dynamics: Design of NMR pulse sequences by gradient ascent algorithms, J. Magn. Reson., 172 (2005) 296-305.
[10]    K. Kobzar, T.E. Skinner, N. Khaneja, S.J. Glaser, B. Luy, Exploring the limits of broadband excitation and inversion: II. Rf-power optimized pulses, J. Magn. Reson., 194 (2008) 58-66.
[11]    Z. Tošner, S.J. Glaser, N. Khaneja, N.C. Nielsen, Effective Hamiltonians by optimal control: Solid-state NMR double-quantum planar and isotropic dipolar recoupling, J. Chem. Phys., 125 (2006) 184502.
[12]    Z. Tošner, T. Vosegaard, C. Kehlet, N. Khaneja, S.J. Glaser, N.C. Nielsen, Optimal control in NMR spectroscopy: Numerical implementation in SIMPSON, J. Magn. Reson., 197 (2009) 120-134.
[13]    N. Khaneja, R. Brockett, S.J. Glaser, Time optimal control in spin systems, Phys. Rev. A, 63 (2001) 323081.
[14]    N. Khaneja, T. Reiss, B. Luy, S.J. Glaser, Optimal control of spin dynamics in the presence of relaxation, J. Magn. Reson., 162 (2003) 311-319.
[15]    T. Calarco, M.A. Cirone, M. Cozzini, A. Negretti, A. Recati, E. Charron, Quantum control theory for decoherence suppression in quantum gates, Int. J. Quant. Inf., 5 (2007) 207-213.
[16]    T. Caneva, M. Murphy, T. Calarco, R. Fazio, S. Montangero, V. Giovannetti, G.E. Santoro, Optimal control at the quantum speed limit, Phys. Rev. Lett., 103 (2009).
[17]    U.V. Poulsen, S. Sklarz, D. Tannor, T. Calarco, Correcting errors in a quantum gate with pushed ions via optimal control, Phys. Rev. A, 82 (2010) 012339.
[18]    N. Khaneja, B. Heitmann, A. Spörl, H. Yuan, T. Schulte-Herbrüggen, S.J. Glaser, Shortest paths for efficient control of indirectly coupled qubits, Phys. Rev. A, 75 (2007) 012322.
[19]    A. Spörl, T. Schulte-Herbrüggen, S.J. Glaser, V. Bergholm, M.J. Storcz, J. Ferber, F.K. Wilhelm, Optimal control of coupled Josephson qubits, Phys. Rev. A, 75 (2007) 012302.
[20]    V.F. Krotov, Quantum system control optimization, Dokl. Math., 78 (2008) 949-952.
[21]    V.F. Krotov, I.N. Fel'dman, Iterative method for solving optimal control problems, Eng. Cyber., 21 (1983) 123-130.

[22]     S. Machnes, U. Sander, S.J. Glaser, P.d. Fouquieres, A. Gruslys, S. Schirmer, T. Schulte-Herbrueggen, Comparing, Optimising and Benchmarking Quantum Control Algorithms in a Unifying Programming Framework, arXiv:1011.4874v2.
[23]     H. Mabuchi, N. Khaneja, Principles and applications of control in quantum systems, Int. J. Rob. Nonl. Con., 15 (2005) 647-667.
[24]     B. Luy, K. Kobzar, T.E. Skinner, N. Khaneja, S.J. Glaser, Construction of universal rotations from point-to-point transformations, J. Magn. Reson., 176 (2005) 179-186.
[25]     K. Kobzar, B. Luy, N. Khaneja, S.J. Glaser, Pattern pulses: Design of arbitrary excitation profiles as a function of pulse amplitude and offset, J. Magn. Reson., 173 (2005) 229-235.
[26]     N. Pomplun, B. Heitmann, N. Khaneja, S.J. Glaser, Optimization of electron-nuclear polarization transfer, Appl. Magn. Reson., 34 (2008) 331-346.
[27]     T.E. Skinner, T.O. Reiss, B. Luy, N. Khaneja, S.J. Glaser, Application of optimal control theory to the design of broadband excitation pulses for high-resolution NMR, J. Magn. Reson., 163 (2003) 8-15.
[28]     R.R. Ernst, G. Bodenhausen, A. Wokaun, Principles of nuclear magnetic resonance in one and two dimensions, Clarendon, 1987.
[29]     F.J. Dyson, The Radiation Theories of Tomonaga, Schwinger, and Feynman, Phys. Rev., 75 (1949) 486-502.
[30]     I. Kuprov, C.T. Rodgers, Derivatives of spin dynamics simulations, J. Chem. Phys., 131 (2009) 234108.
[31]     R.S. Dumont, S. Jain, A. Bain, Simulation of many-spin system dynamics via sparse matrix methodology, J. Chem. Phys., 106 (1997) 5928-5936.
[32]     H.J. Hogben, M. Krzystyniak, G.T.P. Charnock, P.J. Hore, I. Kuprov, Spinach - A software library for simulation of spin dynamics in large spin systems, J. Magn. Reson., 208 (2011) 179-194.
[33]     T.C. Fung, Computation of the matrix exponential and its derivatives by scaling and squaring, Int. J. Numer. Meth. Eng., 59 (2004) 1273-1286.
[34]     G.E. Forsythe, W.R. Wasow, Finite-difference methods for partial differential equations, John Wiley and Sons, New York ; London, 1960.
[35]     J.F. Bonnans, Second-order analysis for control constrained optimal control problems of semilinear elliptic systems, Applied Mathematics and Optimization, 38 (1998) 303-325.
[36]     A.J. Koivo, A second-order variational method for discrete-time optimal control problems, Journal of the Franklin Institute, 286 (1968) 321-330.
[37]     H. Maurer, S. Pickenhain, Second-order sufficient conditions for control problems with mixed control-state constraints, Journal of Optimization Theory and Applications, 86 (1995) 649-667.
[38]     R. Jäntti, S.L. Kim, Second-order power control with asymptotically fast convergence, IEEE Journal on Selected Areas in Communications, 18 (2000) 447-457.
[39]     M. Hinze, K. Kunisch, Second order methods for optimal control of time-dependent fluid flow, SIAM Journal on Control and Optimization, 40 (2002) 925-946.
[40]     R. Fletcher, Practical methods of optimization, 2nd ed. ed., Wiley, 1987.
[41]     W.C. Davidon, Variable Metric Method for Minimization, SIAM J. Opt., 1 (1991) 1-17.
[42]     R.H. Byrd, J. Nocedal, R.B. Schnabel, Representations of quasi-Newton matrices and their use in limited memory methods, Mathematical Programming, 63 (1994) 129-156.
[43]     D.C. Liu, J. Nocedal, On the limited memory BFGS method for large scale optimization, Math. Progr., 45 (1989) 503-528.
[44]     T.F. Coleman, Y. Li, On the convergence of interior-reflective Newton methods for nonlinear minimization subject to bounds, Mathematical Programming, 67 (1994) 189-224.
[45]     T.F. Coleman, Y.Y. Li, An interior trust region approach for nonlinear minimization subject to bounds, Siam Journal on Optimization, 6 (1996) 418-445.

**Figure captions**

**Figure 1**  Taylor approximation accuracy for $|\gamma(z)|$ as a function of approximation order and matrix norm. It should be noted that only power series are in practice affordable in Equation (17) – any rational approximation would require computationally expensive matrix inversions.

**Figure 2**  Numerical convergence profiles in double precision for the Taylor approximation of $\gamma(z)$ as functions of approximation order and matrix norm. The Y axis shows the norm of the difference between the value of $\gamma(z)$ evaluated directly and using the series expansion $\gamma_a(z)$ to a given order. Due to the presence of numerical round-off errors, the scaling and squaring procedure is mandatory for $z > 30$.

**Figure 3**  Norm of the deviation of the finite difference derivative (fourth order central finite difference approximation) from the limit of the commutator series in Equation (19) as a function of the differentiation parameter step.

**Figure 4**  *(Left Panel)* Quality of state transfer as a function of iteration number of the BFGS algorithm (as implemented in Matlab's *fmincon* function [44,45]). The fidelity parameter refers to the quality of magnetization inversion under a 50-point shaped radiofrequency pulse applied to a chain of 31 protons with chemical shifts spread at regular intervals over the range of 8 ppm with strong nearest neighbour *J*-couplings of 20 Hz in a 600 MHz magnet. Pulse duration 5 ms (100 μs per waveform step), pulse amplitude capped at 2500 Hz. State space restriction to three-spin orders involving adjacent spins was used to reduce the matrix dimension involved in the simulation. The starting points in the optimization were set to sequences of uniformly distributed random numbers from the ±1000 Hz interval. The "$k^{th}$ order" labels refer to the number of commutator series terms in Equation (19), "exact" refers to the series that has been summed to machine precision.

*(Right Panel)* **A**: pulse-acquire NMR spectrum of the spin system described above; **B**: magnetization inversion profile under the pulse waveform obtained after 100 iterations with the first-order approximation to the gradient; **C**: magnetization inversion profile obtained after 100 iterations with the "exact" gradient computed using Equation (19).

**Figure 5**  Quality of state transfer as a function of iteration number for three Hessian update schemes as compared to cubic line search steepest descent. DFP stands for Davi-

don-Fletcher-Powell method. The fidelity parameter refers to the quality of magnetization inversion pulse in a 31-spin system as described in the caption to Figure 3. It should be noted that the steepest descent minimization requires 5-10 function evaluations per iteration (line search) and is therefore considerably slower on the wall clock, as well as iteration count, than the three quasi-Newton methods.

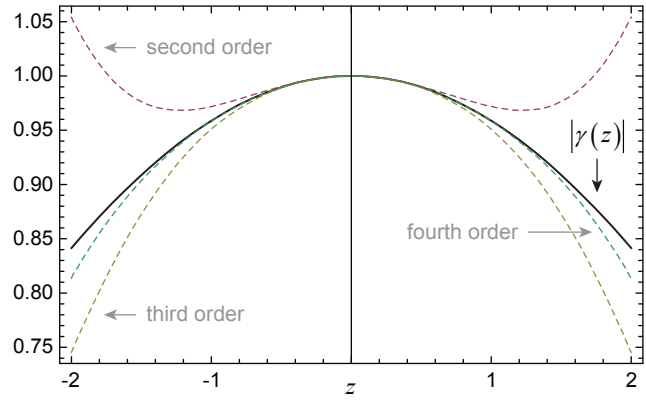

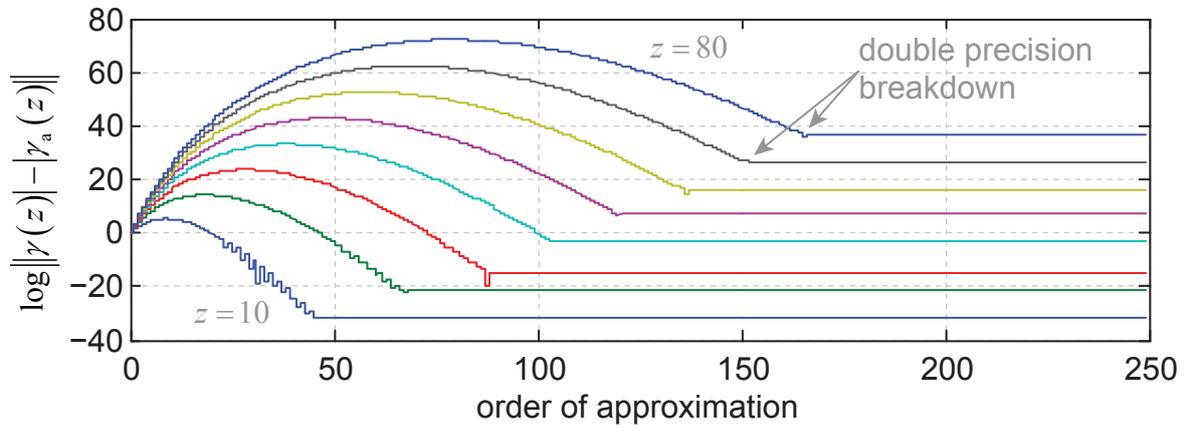

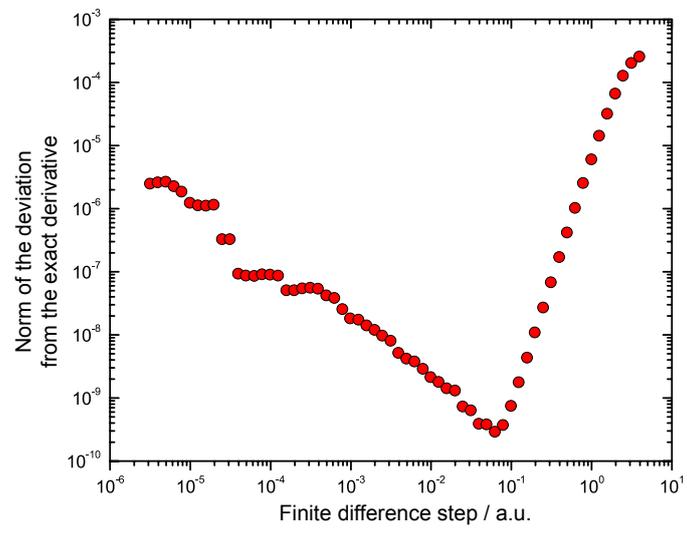

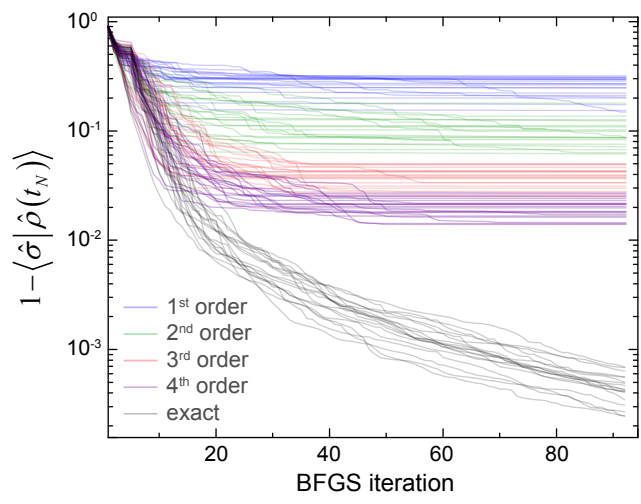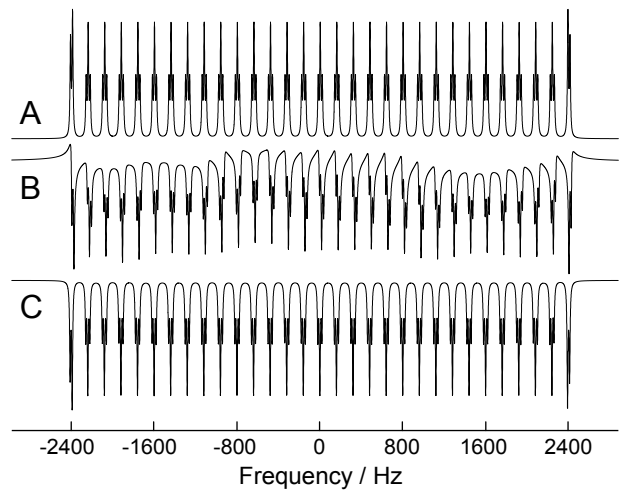

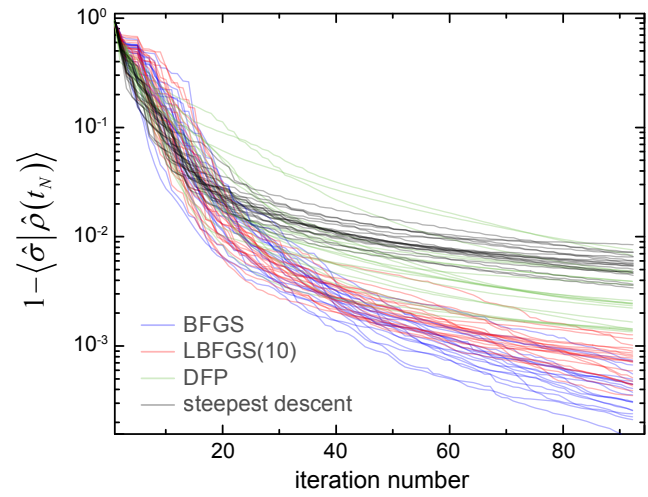